\documentclass[twocolumn,letterpaper]{IEEEAerospaceCLS}  % only supports two-column, letterpaper format

\usepackage[]{graphicx}    % We use this package in this document

\newcommand{\ignore}[1]{}  % {} empty inside = %% comment

\begin{document}
\title{Digital Cellular Solid Pressure Vessels: A Novel Approach for Human Habitation
in Space}

\author{%
Daniel Cellucci\\
Department of Mechanical and Aerospace Engineering\\
Cornell University\\
Ithaca, NY, 14850\\
dwc238@cornell.edu
\and
Benjamin Jenett\\
Center for Bits and Atoms\\
Massachusetts Institute of Technology\\
Cambridge, MA, 02139\\
bej@mit.edu
\and
Kenneth C. Cheung\\
NASA Ames Research Center\\
Moffett Field, CA 94035\\
kenny@nasa.gov
%%%% IMPORTANT: Use the correct copyright information--IEEE, Crown, or U.S. government. %%%%%
%\thanks{\footnotesize 978-1-5090-1613-6/17/$31.00$ \copyright2017 IEEE}              % This creates the copyright info that is the correct 2016 data.
\thanks{{U.S. Government work not protected by U.S. copyright}}         % Use this copyright notice only if you are employed by the U.S. Government.
%\thanks{{978-1-5090-1613-6/17/$\$31.00$ \copyright2017 Crown}}          % Use this copyright notice only if you are employed by a crown government (e.g., Canada, UK, Australia).
%\thanks{{978-1-5090-1613-6/17/$\$310.00$ \copyright2017 European Union}}    % Use this copyright notice is you are employed by the European Union.
}
\maketitle

\thispagestyle{plain}
\pagestyle{plain}

\begin{abstract}
It is widely assumed that human exploration beyond Earth’s orbit will require vehicles capable of providing long-duration habitats that simulate an Earthlike environment - consistent artificial gravity, breathable atmosphere, and sufficient living space- while requiring the minimum possible launch mass. This paper examines how the qualities of digital cellular solids - high-performance, repairability, reconfigurability, tunable mechanical response - allow the accomplishment of long-duration habitat objectives at a fraction of the mass required for traditional structural technologies.

To illustrate the impact digital cellular solids could make as a replacement to conventional habitat subsystems, we compare recent proposed deep space habitat structural systems with a digital cellular solids pressure vessel design that consists of a carbon fiber reinforced polymer (CFRP) digital cellular solid cylindrical framework that is lined with an ultra-high molecular weight polyethylene (UHMWPE) skin. We use the analytical treatment of a linear specific modulus scaling cellular solid to find the minimum mass pressure vessel for a structure and find that, for equivalent habitable volume and appropriate safety factors, the use of digital cellular solids provides clear methods for producing structures that are not only repairable and reconfigurable, but also higher performance than their conventionally-manufactured counterparts.
\end{abstract}

\tableofcontents

%%%%%%%%%%%%%%%%%%%%%%%%%%%%%%%%%%%%%%
\section{Introduction}
%%%%%%%%%%%%%%%%%%%%%%%%%%%%%%%%%%%%%%
Human exploration beyond Earth's orbit will require the creation of structures that can maintain an Earth-like environment with the minimum possible expenditure of mass. One aspect of that environment is atmosphere: that is, being able to create large structural volumes which can contain breathable air at sufficient pressure for human habitation. There are a range of solutions for generating structures which can enclose such volumes under such high pressures without breaching, and this paper examines the application of a building block based approach to the construction of these vessels.

%%%%%%%%%%%%%%%%%%%%%%%%%%%%%%%%%%%%%%
\section{Background}
%%%%%%%%%%%%%%%%%%%%%%%%%%%%%%%%%%%%%%
Cellular solids and architected materials are a recent innovation in material science, allowing the construction of materials with unprecedented specific stiffness and strength\cite{zheng2014ultralight}. Digital cellular solids expand upon the existing work of cellular solids by decomposing the periodic lattices which compose these solids into discrete parts which can then be mass-manufactured and, when combined with a reversible connection, reconfigured and repaired to adapt to changing mission criteria. This approach has been successfully demonstrated in aerospace applications such as shape-morphing aircraft\cite{jenett2015digital} and reconfigurable, mesoscale structures\cite{jenett2016meso}.

%%% NOTE: Add Orion Capsule, modular habs work

Conversely, pressurized habitats have historically been designed as monolithic, thin-shell structures manufactured on Earth that have either been launched as a rigid structure or, more recently, deployed when they reach their intended destination\cite{seedhouse2015bigelow}. There are many proposed habitat types and methods for construction, but two major kinds that have been previously used include inflatables and rigid vessels. Inflatables are launched in a packed configuration and use the atmospheric pressure both as a deployment mechanism and as a means for maintaining the vessel shape. Rigid vessels are composed of a pre-formed rigid material that is launched in its final shape and maneuvered into position.

\subsection{Artificial Gravity Vessels}
Since extended habitation in microgravity environments has been shown to be detrimental to human health\cite{blaber2010bioastronautics}, habitats that can provide an artificial gravity gradient will be necessary for long-term missions. Typically, this gravity is provided by spinning a cylindrical or toroidal vessel such that crew members on the inner surface of the vessel feel a downward centrifugal force. The addition of this spinning movement, however, produces a disorienting Coriolis force, and so the vessel must be of a minimum size in order to reduce this disorienting effect\cite{stone1973overview}. At these minimum sizes, tori require an order of magnitude less mass, since they effectively decouple the atmospheric load from the artificial gravity\cite{o1977space}. 

A toroidal habitat is described with two parameters: the minor radius, or radius of the tube, and the major radius, the radius of the ring that the tube forms. The Stanford Torus is one of the most well-known designs, and proposed a torus with a 65m minor radius and 900m major radius. Subsequent studies have examined improved methods for construction of similar habitats using tensgrity structures\cite{skelton2014growth}.

%%%%%%%%%%%%%%%%%%%%%%%%%%%%%%%%%%%%%%
\section{Methodology}
%%%%%%%%%%%%%%%%%%%%%%%%%%%%%%%%%%%%%%
Comparing the performance of digital material pressure vessels to existing manufacturing approaches requires the selection of equivalent baseline structures that are estimated with similar levels of abstraction assumed. One such structure is the Deep Space Habitat (DSH) described in the Mars Design Reference Architecture (MDRA), Addendum \#2~\cite{drake2010human}.   

The MDRA DSH consists of a 7.2-meter diameter cylinder with ellipsoidal end caps, a total pressurized volume of 290.4 $m^3$, and an atmospheric pressure of 70.3 kPa (0.7 atm) with a safety factor $f_s$ of 2. The total mass of this habitat is given as 5,103 kg, including 300 kg of secondary interior structure, four 0.5 m diameter windows, one exterior hatch, and three docking mechanisms with two docking tunnels~\cite{drake2010human}. These additional components could form a significant proportion of this final mass, and since they are not directly related to the pressure vessel functionality, another method for calculating the expected contribution from the pressure vessel is necessary.

\begin{figure}\label{cylinder}
\centering
\includegraphics[width=3in]{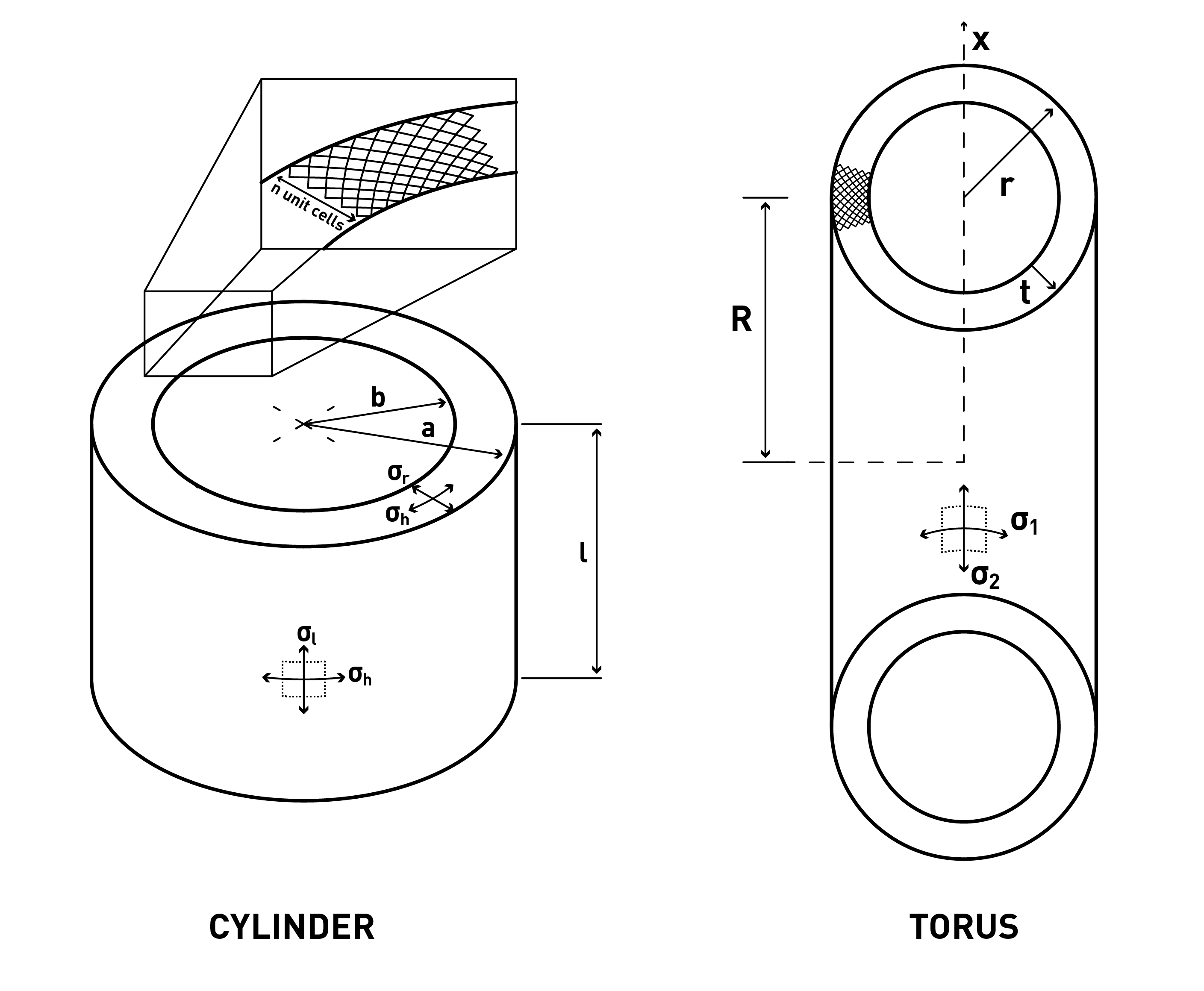}\\
\caption{A cylindrical (left) and toroidal (right) pressure vessel with the important dimensions labelled, as well as the directions of the relevant stresses $\sigma$. Of these, $\sigma_h$ is the most important for the cylinder, and $\sigma_1$ is the most important for the torus. The pressure $P$ is applied to the inner surface of both the cylinder and the torus.}
\end{figure}

\subsection{Rigid Pressure Vessels}

In this vein, Dorsey et. al. provide sizing estimates for pressure vessels in lunar habitats~\cite{dorsey2008structural}, which have been incorporated into the logistical analyses of other missions, including the MDRA DSH~\cite{drake2010human}. Despite a harsh launch environment characterized by significant dynamic and quasi-static loads, the primary load case for a rigid vessel comes from the pressure it contains. Minimum mass estimates are therefore calculated by finding the minimum thickness of a pressure vessel given interior pressure, dimension, and shape. For a cylindrical vessel, this minimum thickness can be found by equating the hoop stress of the cylinder to the material yield stress~\cite{dorsey2008structural}.

\begin{equation}
	t = \frac{f_s P b}{\sigma_Y}
\end{equation}

Where $f_s$ is the factor of safety, $P$ is the pressure, $b$ is the radius of the vessel, and $\sigma_Y$ is the yield stress of the material constituting the vessel.

\subsubsection{Manufacturing Constraints}

Due to the analytical nature of the sizing estimates used here, it is often possible to find a minimum thickness for a pressure vessels that is neither physically realizable nor reliable enough to trust the well-being of a human crew for an extended period of time. As a result, minimum values for the gauge thickness of a pressure vessel for a variety of materials were estimated by the High Speed Research Program~\cite{wilhite2000overview}. These estimates were designed for a primary structure with a 20-year lifetime that meets durability and damage tolerance requirements for commercial transports.

The materials used in this study are Aluminum-Lithium alloy Al-Li 2158-T8, Carbon Fiber Reinforced Polymer (CFRP), and Ultra-high Molecular Weight Polyethylene (UHMWPE). The yield strengths and the minimum gauge thicknesses for these values are given in Table~\ref{MaterialProps}.

\begin{table*}
\renewcommand{\arraystretch}{1.3}
\caption{{\bf Material Properties} for the materials that will be used in this study. Minimum gauge values were given for CFRP and Al-Li, but not for UHMWPE}
\label{MaterialProps}
\centering
\begin{tabular}{|c|c|c|c|c|}
\hline
&\bfseries Al-Li 2541-T8~\cite{dorsey2008structural} & \bfseries CFRP\cite{donald1982adams} & \bfseries UHMWPE (Isotropic)\cite{UHMWPE} & \bfseries Steel\cite{skelton2014growth} \\
\hline\hline
{\bf Yield Strength (MPa)} & 504 & 2500 & 25 & 690\\
{\bf Density ($kg/m^3$)} & 2700 & 1620 & 941 & 7862 \\
{\bf Minimum Gauge (cm)} & 0.102 & 0.203 & - & -\\
\hline
\end{tabular}
\end{table*}

Finally, while it is sufficient to assume that the mechanical behavior of Al-Li and UHMWPE is isotropic, the extreme anisotropy associated with unidirectional CFRP requires special consideration in the design of pressure vessels. The conventional method for manufacturing these vessels using CFRP is to use filament winding, with the fibers oriented 54.7 degrees from the long axis of the cylinder~\cite{sulaiman2013finite}. This alignment ensures that the strength of the resulting composite is twice that of the longitudinal in the hoop direction, and therefore balanced with the expected stresse for the pressurized cylindrical vessel. 

However, this alignment of the fibers comes at the cost of performance relative to the unidirectional material. An FEM study of the effect of the fiber alignment on resulting pressure found that the effective yield strength of the filament-wound vessel was, at best, 28\% that of the unidirectional material~\cite{sulaiman2013finite}. For these analyses, therefore, this corrected ultimate tensile strength will be used in leiu of the unidirectional value. 

\subsection{Digital Pressure Vessels}
Digital cellular solids split the assumption of atmospheric load into two subsystems: a cellular solids structure that maintains integrity and a skin that transfers the atmospheric pressure load to the structure. 

\subsubsection{Cellular Solids Structure}
The cellular solids structure consists of a periodic lattice of struts connected to form a geometry that transfers loads either through bending or axially. The approximate mechanical performance of a cellular solid is proportional to the mechanical performance of the material which constitutes the lattice through the relative density, the ratio of the effective density of the lattice, $\rho^*$, to the density of the constituent material, $\rho$~\cite{deshpande2001effective}. For instance, the ultimate tensile strength of a cellular solid $U^*$ is related to the ultimate tensile strength of its constituent material $U$ by

\begin{equation}
\label{reldenmech}
	U^* = kU\left(\frac{\rho*}{\rho}\right)^a.
\end{equation}

Here, $k$ and $a$ are quantities which depend on the geometry of the cellular solid. For $k$, the value depends on the lattice and the direction of the load. We will use a value of 1/3, the typical coefficient for the anisotropic yield criterion of the octet truss in a principle axis direction~\cite{deshpande2001effective}. This criterion assumes that struts fail through tensile strain before those struts under compressive load fail through buckling. Also for the octet truss, $a$ is equal to 1, since loads are transferred axially rather than through bending. 

Typically, a cellular solid will display behavior approximated by Equation~\ref{reldenmech} when the relative density is below 0.3. Existing work in cellular solids has achieved relative densities less than 0.001 and absolute densities less than 10 $kg/m^3$. This section of the material properties' space is otherwise known as the \emph{ultralight} regime~\cite{zheng2014ultralight}.

The relative density approximation of the mechanical behavior of a cellular solid is subject to one additional requirement: any single dimension of the sample composed of cellular solids must be composed of a minimum number of unit cells of the lattice, in order to ensure that the structure's behavior can be approximated by a continuum. That is, this minimum number of unit cells ensures that the edge effect does not dominate. This minimum number depends on the geometry, but a typical number is 10 cells~\cite{cheung2013reversibly}.
	
The relative density approximation for cellular solids can simplify the analysis of a minimum mass pressure vessel. Since the relative density can relate the material yield stress to the yield stress of the cellular solid, it can act as a design parameter that varies based on the expected stresses experienced by the structure. There are two ways the relative density can be used to approximate the final mass: The first finds the maximum relative density corresponding to the maximum stress in the structure and applies this relative density uniformly throughout, and the second varies the local relative density based on the local conditions of the stress field. While the latter treatment is less realistic than the former, it provides a useful lower bound for examining the effect of having a variable relative density inside a structure has on the overall mass of that structure.

In the case of a thick cylindrical vessel, the primary stress is the hoop stress, which depends on the radial distance $r$ from the center. Using the relationship between relative density $\rho^*/\rho$ and yield stress in Equation~\ref{reldenmech}, and substituting the cellular solids yield stress to the hoop stress of a thick cylinder~\cite{young2002roark}, we find

\begin{equation}
\rho^*(r) = f_s\frac{P}{k}\frac{\rho}{\sigma_Y}\frac{b^2(a^2+r^2)}{r^2(a^2-b^2)}
\end{equation}

The mass of a pressure vessel using the first approach can be found by using the maximum hoop stress in the structure (when $r = b$) to find the required cellular solid density at that point, and then multiplying this density by the volume of the structure to find the total mass. 

\begin{equation}
M_{max} = \pi L f_s\frac{P}{k}\frac{\rho}{\sigma_Y}(a^2+b^2)
\end{equation}

The mass of a pressure vessel using the second approach can be found by finding the $r$-dependent density and integrating it over the volume of the cylinder. The resulting mass is 

\begin{equation}
M_{min} = 2\pi f_sL\frac{P}{k}\frac{\rho}{\sigma}b^2\left(\frac{a^2\textrm{ln}(a/b)}{a^2-b^2}+\frac{1}{2}\right)
\end{equation}

\subsubsection{Digital Pressure Vessel Skin}
The role of the skin in a digital materials pressure vessel is to seal the atmosphere within the structure and transfer the atmospheric load to the frame. Because of the periodic nature of the structure, the panels need only span a single unit cell of the lattice. Since the size of the unit cell above the minimum value does not appreciably affect the material performance, the 'resolution', or number of unit cells in a given volume of structure, becomes another parameter in the design of these vessels.

The octet truss can be oriented relative to the interior surface of the cylinder such that the pattern of struts forms a checkerboard pattern where each square has side length $d/\sqrt{2}$, where $d$ is the length of a unit cell. If rigid skin panel attaches to the entire square perimeter, its minimum thickness can be modeled as the collapse load of a simply supported square plate\cite{young2002roark}, or

\begin{equation}
t = \sqrt{\frac{f_sP}{5.48\sigma_Y}}\frac{d}{\sqrt{2}}.
\end{equation}

This thickness estimates the mass of the panel but not the mass of the interface between the panel and the structure. If this mass is assumed to equal the mass of the structure the panel is attaching to, then the first order approximation of the strut cross-sectional area at the interface is necessary. Assuming cylindrical struts:

\begin{equation}
\frac{\rho^*}{\rho} = 12\sqrt{2}\pi\left(\frac{u}{d}\right)^2 = f_s\frac{P}{k}\frac{1}{\sigma_{Yc}}\left(\frac{a^2-b^2}{a^2+b^2}\right)
\end{equation}

The area $\pi u^2$ is then multiplied by the perimeter of the panel to find the volume of the structure. The total skin mass can then be found by taking the mass of the panel and the attaching structure and multiplying by the number of panels, which can be found by dividing the interior surface area by the area of a single panel. That is,

\begin{equation}
M_{skin} = 2\pi b(L+2b)d \left[\frac{\rho_u}{\sqrt{5.48\sigma_{Yu}}}\sqrt{\frac{f_s P}{2}} + \frac{1}{3}\frac{f_s P}{k}\frac{\rho_c}{\sigma_{Yc}}\frac{a^2-b^2}{a^2+b^2}\right]
\end{equation}

Where $\sigma_{Yu}$ and $\rho_u$ are the yield strength and density of UHMWPE and $\sigma_{Yc}$ and $\rho_c$ are the yield strength and density of CFRP, respectively. 

Examining the relative effect of these two terms, we can see that the first term contributes $\approx$ 200 times to the mass of the structure than the second term. As a result, we can conclude that the mass of the skin is dominated by the panels and not by the interface between the panel and the structure.

\subsubsection{Toroidal Vessels}
Digital material toroidal vessels can be characterized by three parameters: the minor radius $r$, major radius $R$, and thickness $t$. Finding the optimal mass for a toroidal pressure vessel composed of digital cellular solids has the same approach as the cylindrical vessel analysis- either a uniform relative density structure that is rated for the highest stress (upper-bound), or a graded relative-density structure that is rated for only the local stress field (lower-bound). Both approaches rely on the local stress $\sigma_1(x)$, or

\begin{equation}
\frac{Pr}{2t}\frac{x+R}{x}
\end{equation}
Where $x$ is the distance from the center of the major axis and $t$ is assumed to be less than $0.1 r$\cite{young2002roark}

For the upper-bound mass, the highest stress in the structure is located at $x = R-r$. There, 
\begin{equation}
\sigma_1 = \frac{Pr}{2t}\frac{2R-r}{R-r}
\end{equation}

and 

\begin{equation}
\rho^*(R-r) = \frac{1}{k}\frac{\rho}{\sigma}\frac{Pr}{2t}\frac{2R-r}{R-r}
\end{equation}

so the total mass is

\begin{equation}
M_{max} = \frac{\pi^2}{k}\frac{\rho}{\sigma}PRr\left(2r-t\right)\frac{2R-r}{R-r}
\end{equation}

For the lower-bound mass, the space-varying relative-density is necessary

\begin{equation}
\rho^*(x) = \frac{1}{k}\frac{\rho}{\sigma}\frac{Pr}{2t}\frac{x+R}{x}
\end{equation}

Which, multiplying by the volume and integrating is the total volume of the structure $M$

\begin{equation}
M_{min} = \int{(2\pi x)(2 h(x)) \rho^*(x)\textrm{d}x}
  = \frac{2\pi}{k}\frac{\rho}{\sigma}\frac{Pr}{t}\int{(x+R)h(x)}
\end{equation}

where $h(x)$ is the area of the cross-section of the torus. Integrating $x$ from $R-r$ to $R+r$ produces the equation for the total mass

\begin{equation}
M_{min} = 2\pi^2\frac{1}{k}\frac{\rho}{\sigma}PRr(2r-t)
\end{equation}

The ratio of the upper-bound mass to the lower-mass is then:

\begin{equation}
\frac{M_{max}}{M_{min}} = \frac{2R-r}{2(R-r)}
\end{equation}

%%%%%%%%%%%%%%%%%%%%%%%%%%%%%%%%%%%%%%
\section{Results}
%%%%%%%%%%%%%%%%%%%%%%%%%%%%%%%%%%%%%%
Applying the equations above allows the comparison of the total mass of the rigid monolithic pressure vessel to that of the digital. For these monolithic vessels, the minimum masses are shown in Table~\ref{RigidMasses}. In the case of the Al-Li vessel, the final size of the vessel was larger than the minimum gauge, but in the case of the CFRP filament-wound vessel the minimum thickness was half the minimum gauge, resulting in a structure with twice the mass of the version unconstrained by the minimum gauge requirement. 

\begin{table}
\renewcommand{\arraystretch}{1.3}
\caption{{\bf Monolithic Pressure Vessel Minimum Mass} for a Mars Design Reference Architecture Deep Space Habitat. Note that the CFRP vessel mass is determined by the minimum allowable gauge, and not the theoretical minimum thickness}
\label{RigidMasses}
\centering
\begin{tabular}{|c|c|}
\hline
\bfseries Material &\bfseries Mass (kg)  \\
\hline\hline
{\bf Al-Li 2541-T8} & 833 \\
{\bf CFRP} &  709 \\
{\bf UHMWPE} & 4122 \\
\hline
\end{tabular}
\end{table}

Two additional degrees of freedom, the overall thickness of the vessel and the pitch of the lattice that composes the structural subsystem, are available in the design of a digital materials pressure vessel.

\begin{figure*}\label{massfigs}
\centering
\includegraphics[width=6in]{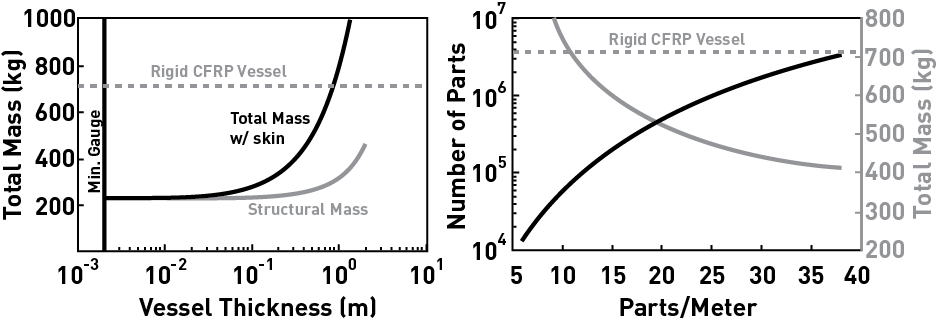}\\
\caption{\textbf{Comparison of a rigid CFRP and Digital CFRP/UHMWPE pressure vessel. The left plot shows the total mass of a vessel with changing thickness, assuming that the unit cell of the structure is one-tenth the thickness. The right plot shows the effect of decreasing this unit cell size on the total mass of the digital pressure vessel and the overall complexity of the structure.}} 
\end{figure*}

\subsection{Shell Thickness}
The left chart in Figure~\ref{massfigs} shows the structural mass and the total mass for a digital MDRA DSH pressure vessel whose thickness varies from the minimum gauge of a rigid CFRP pressure vessel, 0.204 cm, to 2 meters. The total mass assumes that the unit cell is 1/10th the thickness of the vessel in order to estimate the mass of the skin.

\subsection{Lattice Pitch}
The right chart in Figure~\ref{massfigs} shows the total mass of a MDRA DSH with a 1m thick hull as the number of parts per meter is increased, as well as the total number of parts in the structure. The structural mass is relatively unchanged while the resolution increases, remaining at 292 kg.

\subsection{Toroidal Vessel}
Using an equivalent approach to the one descibed for the cylindrical vessel, a digital toroidal pressure vessel with a major radius of 10m, a minor radius of 5m, a thickness of 0.5m, and a lattice pitch of 31mm has a structural mass between 2606 and 3518 kg. 

\subsubsection{Comparison to Other Proposed Designs}
This result compares favorably to other proposed designs, such as the cable network inflatables of Hoyt et. al.\cite{skelton2014growth} There, the authors used the unidirectional strength of UHMWPE for both the skin of the structure as well as the cable network that reinforces the skin. Since the skin is assumed to be a rigid dome, the material properties for an isotropic sheet such as those listed in Table~\ref{MaterialProps} provide a more accurate measure of the skin mass and therefore the total mass of the system. As a result of this substitution, the total mass of the cable network torus is 3490kg, which is within the best-case and worst-case limits for the ideal digital cellular solids design.

%%%%%%%%%%%%%%%%%%%%%%%%%%%%%%%%%%%%%%
\section{Discussion}
\begin{figure}\label{partsizes}
\centering
\includegraphics[width=3in]{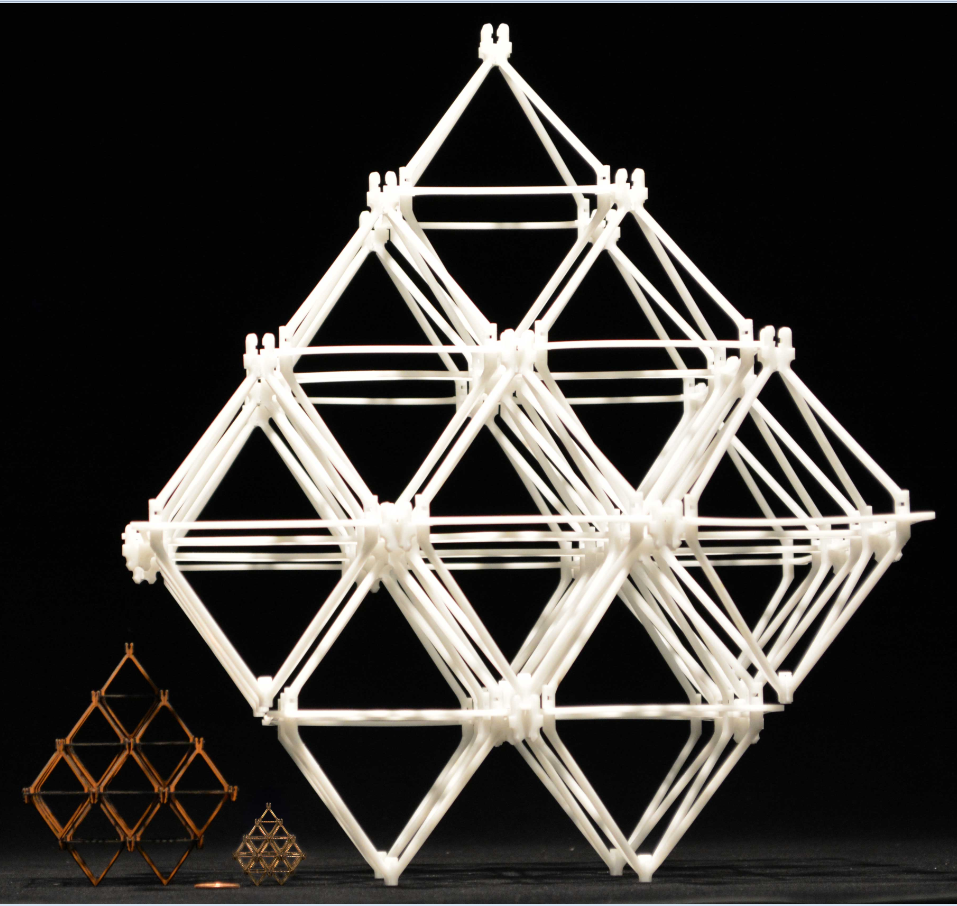}\\
\caption{Three Different Part Sizes for three different levels of resolution. Digital cellular solids can be manufactured at a variety of length scales.}
\end{figure}

%%%%%%%%%%%%%%%%%%%%%%%%%%%%%%%%%%%%%%
The two charts in Figure~\ref{massfigs} illustrate a few important points in the design of digital pressure vessels. First, the vessel thickness can vary almost three orders of magnitude without resulting in an appreciable change in the overall mass of the structure. Since a thicker vessel can use larger parts, it is therefore easier to assemble. However, and second, these larger parts come at a cost, since the mass of the skin can quickly out weigh that of the structure if the parts become too large and the space the panels have to bridge become too wide. The skin mass can be reduced without dramatically increasing the structural mass by simply increasing the resolution of the structure. However, this increased resolution comes at the cost of complexity, since the number of parts in a structure is proportional to $d^3$ while the skin mass is proportional to only $d$. Figure~\ref{partsizes} shows the octet structure composed of parts at three different length scales. Smaller scales produce higher performance but are also harder to manufacture and assemble. 

\subsection{Growth}
\begin{figure*}\label{growth}
\centering
\includegraphics[width=6in]{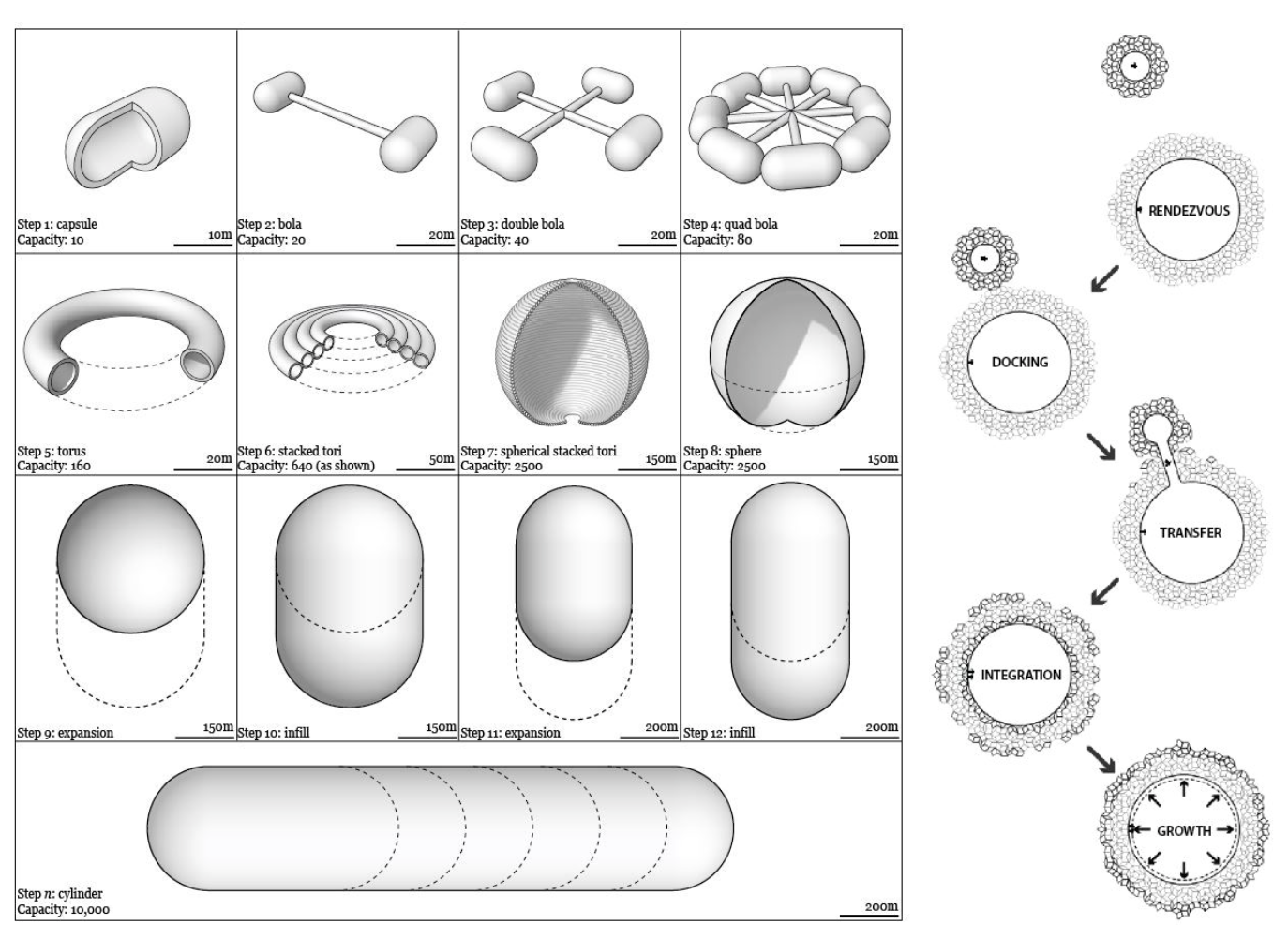}\\
\caption{The left figure shows different size structures, from a ten-person capsule to a 10,000 person colony, and the morphologies which represent the most efficient configuration of mass at each scale\protect\cite{johnson1977space}. The right figure shows a schematic representation of how a structure composed of digital cellular solids could transition from one morphology to another.}
\end{figure*}

The reconfigurability shown in previous work~\cite{jenett2016meso} suggests that the use of digital cellular solids allows for flexibility regarding the design of a habitat system. Namely, the size and shape of the habitat can change based on mission criteria in order to adapt to the current conditions. Figure~\ref{growth} shows a succession of habitats of increasing size, ranging from a capacity of 10 to 10,000 humans, as well as a schematic description of how two digital cellular solids structures could rendezvous and consolidate their parts into a larger structure.

%%%%%%%%%%%%%%%%%%%%%%%%%%%%%%%%%%%%%%
\section{Conclusion}
%%%%%%%%%%%%%%%%%%%%%%%%%%%%%%%%%%%%%%
It is clear from this analysis that digital cellular solids can confer improved performance over conventionally-manufactured rigid pressure vessels. Additionally, examination of toroidal pressure vessels shows that the digital cellular solids can compete with other proposed habitat construction designs without sacrificing the core capabilities that allow it to grow and adapt.

%%%%%%%%%%%%%%%%%%%%%%%%%%%%%%%%%%%%%%%%%%%%%%%%%%%%%%%%%%%%%%%%%%%%%%%%%%%%%%%%%%%%%%%%%%%%%%%%%%%%%%
\acknowledgments
The authors thank the Ames Center Innovation Fund for supporting this project. This project was also supported by the NASA Space Technology Research Fellowship, Grants \#NNX13AL38AH and \#NNX14AM40H.

%%%%%%%%%%%%%%%%%%%%%%%%%%%%%%%%%%%%%%%%%%%%%%%%%%%%%%%%%%%%%%%%%%%%%%%%%%%%%%%%%%%%%%%%%%%%%%%%%%%%%%
\bibliographystyle{IEEEtran}
\bibliography{mybibfile}

%%%%%%%%%%%%%%%%%%%%%%%%%%%%%%%%%%%%%%%%%%%%%%%%%%%%%%%%%%%%%%%%%%%%%%%%%%%%%%%%%%%%%%%%%%%%%%%%%%%%%%
\thebiography
%% This biostyle allows you to insert your photo size 1in X 1.25in
\begin{biographywithpic}
{Daniel Cellucci}{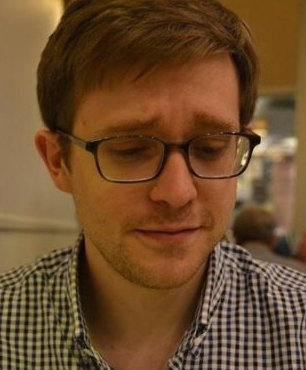}
received his B.S. degree in Physics from the University of Georgia in 2012, and is currently a Ph.D. candidate in Mechanical Engineering at Cornell University. He is a NASA Space Technology Research Fellow and currently studies the robotic assembly of large-scale space structures.
\end{biographywithpic} 

\begin{biographywithpic}
{Benjamin Jenett}{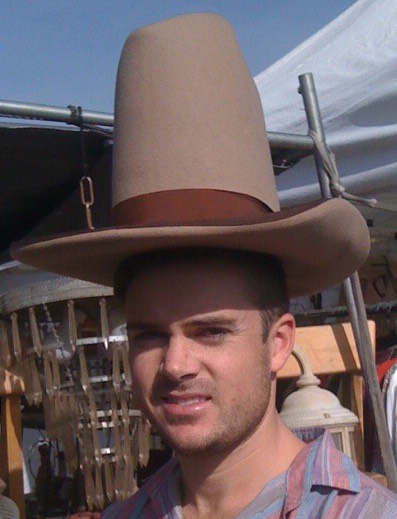}
is a graduate student researcher at MIT’s Center for Bits and Atoms, where he is pursuing his Ph.D. studying automated assembly for large aerospace structures. He is a NASA Space Technology Research Fellow.

\end{biographywithpic}

\end{document}